\documentclass[12pt]{iopart}
\usepackage {graphicx}
\usepackage{iopams}
\usepackage{color}

\begin{document}

\def\be{\begin{equation}}
\def\ee{\end{equation}}
\def\correct{}

\title[Non-relativistic limits of Maxwell's equations]{Non-relativistic limits of Maxwell's equations}
\author{Giovanni Manfredi}
\address{Institut de
Physique et Chimie des Mat\'{e}riaux, CNRS and Universit\'{e} de
Strasbourg, BP 43, F-67034 Strasbourg, France}
\ead{giovanni.manfredi@ipcms.unistra.fr}

\date{\today}

\begin{abstract}
In 1973, Le Bellac and L\'{e}vy-Leblond ({\it Nuovo Cimento} B {\bf 14} 217--234) discovered that Maxwell's equations possess two non-relativistic Galilei-covariant limits, corresponding  to $| {\mathbf E}| \gg c|{\mathbf B}|$ (electric limit) or $| {\mathbf E}| \ll c|{\mathbf B}|$ (magnetic limit). Here, we provide a systematic, yet simple, derivation of these two limits based on a dimensionless form of Maxwell's equations and an expansion of the electric and magnetic fields in a power series of some small parameters. Using this procedure, all previously known results are recovered in a natural and unambiguous way. Some further extensions are also proposed.

\end{abstract}

\maketitle

\section{Introduction}\label{sec:intro}
Maxwell's equations were the first physical theory to be put forward that is fully Lorentz covariant, well before the special theory of relativity was developed and understood. Indeed, it was the clash between the Lorentz covariance of Maxwell's equations and the Galilei covariance of Newtonian mechanics that stimulated the discovery of special relativity and revealed that Lorentz covariance is the most fundamental symmetry of the two.

Nevertheless, it should be stressed that electricity and magnetism were developed historically as independent phenomena and only lately were realized to be different expressions of a single underlying theory. Already in the eighteenth century, physicists like Charles-Augustin Coulomb (1736--1806) manipulated electric charges and measured how they attract or repel each other through electric fields -- in modern language, they worked out applications of Gauss's law: $\nabla \cdot {\mathbf E} = \rho/\varepsilon_0$. This is the field of electrostatic.

One century later, physicists working on magnetism manipulated currents and measured how they interact with each other through magnetic fields.
Around 1820, Oersted observed that wires carrying electric currents deflected a compass needle placed in their vicinity. Biot and Savart, and later Amp\`{e}re, established rigorous laws that related the strength and direction of a magnetic field to the currents that produce it.
In modern language, they worked out applications of Amp\`{e}re's law: $\nabla \times \mathbf{B} =\mu_0 \mathbf{J}$. This is the field of magnetostatics.

But none of these phenomena involve anything that is ``relativistic". This is obvious for electrostatics, as only electric charges and fields are involved; but it is true for magnetostatics too, because the current in Amp\`{e}re's law does not need to be relativistic in order to generate a finite magnetic field.

Thus, we are faced with two perfectly valid theories that appear to be both non-relativistic and both deriving from the complete theory of electromagnetism, i.e. Maxwell's equations. But surely any relativistic theory should have a unique non-relativistic limit -- or should it?

Enter Le Bellac and L\'{e}vy-Leblond (hereafter, LBLL), who in 1973 published a paper where it was demonstrated that Maxwell's equations possess {\em two} distinct, yet perfectly consistent, non-relativistic limits \cite{LBLL}. The two limits correspond to situations where either $|{\mathbf E}| \gg c|{\mathbf B}|$ (electric limit) or $|{\mathbf E}| \ll c|{\mathbf B}|$ (magnetic limit). Each of the two limits is Galilei covariant, although the transformations of the fields and of the density and current are not the same in the two cases. In practice, the electric limit amounts to neglecting the time-derivative of the magnetic field in Faraday's law of induction, whereas the magnetic limit is obtained by dropping the displacement current in Amp\`{e}re's equation.

LBLL's paper basically contains all that one needs to know on this topic, but {\correct has also} several drawbacks. First, the results were presented without a systematic derivation; only {\it a posteriori} it was checked that both limits are indeed Galilei covariant. Second, the authors obtained their result by employing SI units, which was sort of a novelty at their time of writing, CGS Gaussian units {\correct then being} the preferred choice. But any physically meaningful result should not depend on the units in which the equations are written, and a proper limit should be obtained by making use of dimensionless quantities.

More recently, several papers revisited the work of LBLL from different angles, elucidating some issues such as gauge invariance \cite{Rousseaux} and the correct Galilean limit of the relativistic four-current \cite{Holland}, discussing applications \cite{Montigny2007}, or else extending the analysis to the coupled Dirac-Maxwell equations \cite{Holland-Brown}. Some papers use more abstract methods that rely on a five-dimensional Minkowski manifold \cite{Santos,Montigny2003}. A recent work by Degond et al. was devoted to the analysis of a numerical scheme for the Euler-Maxwell equations in the magnetic limit \cite{Degond}.

In none of these recent works the electric and magnetic limits were derived systematically from the full Maxwell equations.
In general, the relations $|{\mathbf E}| \gg c|{\mathbf B}|$ or $|{\mathbf E}| \ll c|{\mathbf B}|$ (and the analogue expressions for the sources) are assumed {\it ex nihilo} and are used to derive the low-velocity transformations of the fields and sources. It is then proven that some form of ``reduced" Maxwell equations are Galilei invariant under those transformations \cite{Holland-Brown}.
Only in one case this analysis is carried out using dimensionless variables \cite{Heras2010} (see also \cite{Degond} and \cite{Heras2009}). Sometimes, non-systematic ``order of magnitude" arguments are used heuristically to justify the limits \cite{Montigny2006}.

The purpose of the present paper is to introduce a simple, yet systematic, procedure to recover the non-relativistic limits of Maxwell's equation. This procedure is based on Maxwell's equations written in a non-dimensional form. It is shown that two independent dimensionless parameters naturally appear in the equations. The electric and magnetic limits are then {\em derived} by letting either or both these parameters go to zero. Using the same procedure, all known properties (gauge relations, Lorentz transformations, the Lorentz force) of the non-relativistic limits can be deduced systematically.
This procedure is also better adapted to possible extensions of the present work, such as the coupling of the Maxwell and Dirac equations in relativistic mean field theories \cite{Masmoudi, Sulaksono}.

\section{General procedure}\label{sec:general}
We start from Maxwell's equations in SI units:
\begin{eqnarray}
\nabla\cdot {\mathbf E} &=& {\rho \over \epsilon_0},  \\
\nabla\cdot {\mathbf B} &=& 0,  \\
\nabla\times {\mathbf E} &=& - \frac{\partial {\mathbf B}}{\partial t}, \\
\nabla\times {\mathbf B} &=& \mu_0 {\mathbf J} + \frac{1}{c^2}\frac{\partial {\mathbf E}}{\partial t}
\,, \label{maxwell}
\end{eqnarray}
where $\varepsilon_0$ and $\mu_0$ are respectively the electric permittivity and the magnetic permeability in vacuum, and $\varepsilon_0 \mu_0 = c^{-2}$.

We normalize space to a reference length $L$ and time to a reference time $T$, which define a typical velocity $V=L/T$. The fields and the sources are also normalized to reference quantities denoted by an overline: $\overline{E}$, $\overline{B}$, $\overline{\rho}$, and $\overline{J}$, satisfying the relations: $\overline{E}= c \bar B$ and $\overline{J} = V \overline{\rho}$.

In these units, Maxwell's equations can be rewritten as:
\begin{eqnarray}
\nabla\cdot {\mathbf E} &=& \frac{\rho}{\alpha}  \\
\nabla\cdot {\mathbf B} &=& 0  \\
\nabla\times {\mathbf E} &=& - \beta \frac{\partial {\mathbf B}}{\partial t} \\
\nabla\times {\mathbf B} &=& \frac{\beta}{\alpha} {\mathbf J} + \beta\frac{\partial {\mathbf E}}{\partial t}
\,,
\label{maxwell_norm}
\end{eqnarray}
where the following two dimensionless parameters have appeared naturally:
\be
 \beta = \frac{V}{c}~, ~~~~~~~ \alpha = \frac{\overline{E} \epsilon_0}{ \overline{\rho} L}~.
\ee
The first of these parameters, $\beta$, is just the reference velocity normalized to the speed of light and is obviously small in any non-relativistic limit.

The parameter $\alpha$ requires a little analysis to
be expressed in more familiar terms. We can think of our distribution of charges and currents as a classical plasma, with a certain temperature $T_0$ and typical charge density $\overline{\rho}$. We then define a reference electrostatic potential $\overline{\phi} = \overline{E}L$ and express it in terms of the temperature, such that $q\overline{\phi}=k_B T_0$, where $q$ is the electric charge of the particles and $k_B$ is the Boltzmann constant.
Then $\alpha$ can be written as:
\be
\alpha = \left(\frac{\lambda_D}{L}\right)^2 \label{alpha}
\ee
where $\lambda_D=\sqrt{k_B T_0 \varepsilon_0/(q\bar \rho)}$ is the Debye length.
In plasma physics, this ratio is known as the ``quasi-neutrality" parameter  and it is small when deviations from local charge neutrality are negligible.
This is because, in most plasmas, charge imbalance can persist only on length scales shorter than the Debye length \cite{Chen, Joyce}. If the latter is very small, the plasma is almost neutral at macroscopic scales $\sim L$.
This requires, of course, that at least two types of charges, positive and negative, be present in the system under consideration.

In the forthcoming sections we will show that the electric and magnetic limits can be obtained by making suitable assumptions on the parameters $\alpha$ and $\beta$. In particular:
\begin{itemize}
\item If $\beta \ll 1$ and $\alpha = O(1)$, we obtain the {\it electric limit};
\item If $\beta \ll 1$ and $\alpha \ll 1$, but $\alpha/\beta = O(1)$, we obtain the {\it magnetic limit}.
\end{itemize}

\section{Electric limit}\label{sec:electric_limit}
For convenience of notation, we define a smallness parameter $\epsilon \ll 1$.
Then, for the electric limit, we assume $\beta \sim \epsilon$ and $\alpha = O(1)$. We expands both fields in a power series in $\epsilon$, such that: ${\mathbf E}={\mathbf E}_0 + \epsilon {\mathbf E}_1 + \dots$ and ${\mathbf B}={\mathbf B}_0 + \epsilon {\mathbf B}_1 + \dots$, where $\mathbf{E}_0$, $\mathbf{E}_1, ~\rm etc \dots$ are quantities of order unity. The density and current are also assumed to be of order unity.

To lowest (zeroth) order in $\epsilon$ one obtains
\begin{eqnarray}
\nabla\cdot {\mathbf E}_0 &=& \rho/\alpha, \label{order0-1}\\
\nabla\cdot {\mathbf B}_0 &=& \nabla\times {\mathbf E}_0 = \nabla\times {\mathbf B}_0=0.
\label{order0}
\end{eqnarray}
Thus $\mathbf{B}_0=0$ \footnote{Strictly speaking, the fact that $\nabla\cdot {\mathbf B}_0 = \nabla\times {\mathbf B}_0=0$ implies $\Delta \mathbf{B}_0=0$, but if $\mathbf{B}_0$ vanishes at infinity, then it must be zero everywhere. Similar considerations apply to vector fields with zero curl and divergence encountered later in this paper.} and we can write $\mathbf{E}_0 = -\nabla \phi_0$, so that:
\be
\Delta \phi_0 = -\rho/\alpha
\label{poisson}
\ee

Let us now proceed to first order. We find
\begin{eqnarray}
 \nabla \cdot \mathbf{E}_1 &=& \nabla \cdot \mathbf{B}_1 = \nabla \times \mathbf{E}_1 = 0, \\
 \nabla \times \mathbf{B}_1 &=& \left(\frac{\mathbf J}{\alpha} + \frac{\partial \mathbf{E}_0}{\partial t}\right) ,
 \label{order1}
\end{eqnarray}
which imply that $\mathbf{E}_1=0$. Introducing the vector potential at first order, ${\mathbf B}_1 =\nabla \times \mathbf{A}_1$, and substituting into Eq. (\ref{order1}), we obtain
\be
\nabla(\nabla \cdot \mathbf{A}_1) - \Delta \mathbf{A}_1 = \mathbf{J}/\alpha - \partial_t(\nabla \phi_0),
\ee
which reduces to
\be
\Delta \mathbf{A}_1 = - \mathbf{J}/\alpha
\label{order1A}
\ee
if we adopt the Lorentz gauge relation
\be \frac{\partial \phi_0}{\partial t} + \nabla
\cdot \mathbf{A}_1 = 0~.
\label{gauge}
\ee

Putting together the results at zeroth and first order, Maxwell's equations in the electric limit can be written in terms of the potentials
\begin{eqnarray}
\Delta \phi_0 &=& - \rho/\alpha, \label{elect_lim_pot_a}\\
\Delta \mathbf{A}_1 &=& - \mathbf{J}/\alpha, \label{elect_lim_pot_b}
\end{eqnarray}
or in terms of the fields
\begin{eqnarray}
\nabla\cdot \mathbf{E}_0 &=& \rho/\alpha, \label{elect_lim_fields1} \\
\nabla\cdot \mathbf{B}_1 &=& \nabla \times \mathbf{E}_0 =0,  \\
\nabla \times \mathbf{B}_1 &=& \frac{\mathbf J}{\alpha} + \frac{\partial \mathbf{E}_0}{\partial t}\,.
\label{elect_lim_fields3}
\end{eqnarray}
Note that only $\mathbf{E}_0$ and $\mathbf{B}_1$ appear in the above equations, because $\mathbf{E}_1=\mathbf{B}_0=0$.
Equations (\ref{elect_lim_fields1})-(\ref{elect_lim_fields3})
are identical to the equations postulated by LBLL for the electric limit \cite{LBLL}. As anticipated in the introduction, they can be heuristically obtained from the full Maxwell's equations by dropping Faraday's induction term. Here the same result was derived from a systematic expansion in powers of a small parameter. The present method also allowed us to recognize that the electric and magnetic fields actually appear at different orders in $\epsilon$.

\paragraph{Lorentz force.---}
Let us now evaluate the Lorentz force per unit volume: $\delta \mathbf{F}= \rho\mathbf{E}+\mathbf{J}\times \mathbf{B}$.
In our units, and using as a reference value for the force $\overline{\delta F} = \overline{\rho} \overline{E}$, we obtain
\be
\delta\mathbf{F}= \rho\mathbf{E}+ \beta \mathbf{J} \times \mathbf{B},
\label{lorentz_force}
\ee
where it appears that the magnetic term is of higher order. Since in the electric limit only $\mathbf{E}_0$ and $\mathbf{B}_1$ are non-vanishing and $\beta \sim \epsilon$, we get
\be
\delta \mathbf{F}= \rho\mathbf{E}_0 + \epsilon^2 \mathbf{J} \times \mathbf{B}_1.
\ee
Thus we conclude that first order terms in Maxwell's equations
will induce a {\em second} order magnetic correction in the particle dynamics, which can be neglected in the present approximation.

\paragraph{Condition of validity.---}
Next, we would like to derive the condition of validity for the electric limit as established by LBLL, i.e., $|\mathbf{E}| \gg c|\mathbf{B}|$. For this purpose, we compute the relative strength of the electric and magnetic fields. One can write
\be
\frac{|\mathbf{E}|}{|\mathbf{B}|} \approx \frac{\overline{E}}{\overline{B}} ~\frac{E_0 + \epsilon E_1 + \dots}{B_0 + \epsilon B_1 + \dots} = c~\frac{E_0 + \epsilon E_1 + \dots}{B_0 + \epsilon B_1 + \dots}.
\label{ratio_EB}
\ee
For the electric limit ${E}_1={B}_0 = 0$, whereas $E_0 \sim B_1 = O(1)$. Therefore we obtain:
\be
\frac{|\mathbf{E}|}{|\mathbf{B}|} \approx c~\frac{{E}_0}{\epsilon {B}_1} \approx \frac{c}{\epsilon},
\ee
and since $\epsilon \ll 1$, we find that $|\mathbf{E}| \gg c|\mathbf{B}|$,
which is the expected result.

Finally, the continuity equation can be obtained by taking the divergence of Eq. (\ref{order1}) and using Eq. (\ref{order0-1}), which yields
\be
\frac{\partial \rho}{\partial t} + \nabla\cdot \mathbf{J} = 0
\label{continuity}
\ee

\section{Magnetic limit}\label{sec:magnetic_limit}
This is obtained by taking $\beta \to 0$ and $\alpha \to 0$, but keeping
the ratio $\beta/\alpha$ finite. In other words $\alpha \sim \beta \sim \epsilon \ll 1$.
By performing the same expansion as in Sec. \ref{sec:electric_limit},
we obtain at zeroth order
\begin{eqnarray}
\nabla\cdot \mathbf{E}_0 &=& \nabla\times \mathbf{E}_0 = \nabla\cdot \mathbf{B}_0=0,
\\
\nabla\times \mathbf{B}_0 &=&  \mathbf{J}
\,.
\label{order0_mag}
\end{eqnarray}
Note that Gauss's law also implies that $\rho=0$ if we assume -- as we did so far -- that the charge density is a zeroth order quantity (a more general case will be discussed shortly). Thus, the magnetic limit deals with systems that are locally charge neutral, a fact that was already acknowledged by LBLL.

Equation (\ref{order0_mag}) represents the {\it magnetostatic} limit: no free charges, only currents; no electric fields, only magnetic fields. We also note that the current is divergence free.

If we pursue the expansion to first order, we obtain
\begin{eqnarray}
\nabla \times \mathbf{B}_1 &=& \nabla \cdot \mathbf{B}_1=\nabla\cdot \mathbf{E}_1 = 0  \label{order1_mag1}\\
\nabla\times \mathbf{E}_1 &=& -\frac{\partial \mathbf B_0}{\partial t}.
\label{order1_mag2}
\end{eqnarray}
Summarizing the results at zeroth and first order, we can write:
\begin{eqnarray}
\nabla \cdot \mathbf{B}_0 &=& \nabla \cdot \mathbf{E}_1 = 0 \label{order01_mag1}\\
\nabla\times \mathbf{B}_0 &=&  \mathbf{J} \label{order01_mag2}\\
\nabla\times \mathbf{E}_1 &=& -\frac{\partial \mathbf{B}_0}{\partial t},
\label{order01_mag3}
\end{eqnarray}
and $\mathbf{B}_1 = \mathbf{E}_0 =0$.
We note that this is the approximation of Maxwell's equations used in magneto-hydrodynamics (MHD). Indeed, MHD is a one-fluid theory that describes the motion of a fluid that carries electric currents but no electric charge. The currents generate a self-consistent magnetic field through Eq. (\ref{order01_mag2}), which in turn acts back on the fluid via the Lorentz force.
Equations (\ref{order01_mag1})-(\ref{order01_mag3}) are almost identical to the equations postulated by LBLL for the magnetic limit \cite{LBLL}, except that those authors found $\nabla\cdot \mathbf{E}_1 \neq 0$. We will return on this point later.

Introducing the vector potential $\mathbf{B}_0 = \nabla \times \mathbf{A}_0$ and using the Coulomb gauge $\nabla\cdot \mathbf{A}_0 =0$, yields
\be
\Delta \mathbf{A}_0 =-\mathbf{J}.
\ee
Thus, it appears that the Coulomb gauge is the natural choice for the magnetic limit.

In dimensionless units, the electric field is written in terms of the potentials:
\be
 \mathbf{E} = -\nabla \phi - \epsilon \frac{\partial \mathbf A}{\partial t}.
 \label{efield}
\ee
At first order, this becomes
\be
 \mathbf{E}_1 = -\nabla \phi_1 - \frac{\partial \mathbf A_{0}}{\partial t},
 \label{efield1}
\ee
which satisfies automatically Eq. (\ref{order1_mag2}).

\paragraph{Charge neutrality.---}
The charge neutrality condition ($\rho=0$) is correct only if one assumes that the density is a zeroth-order quantity in $\epsilon$. Different equations are obtained if the density is a first or second order quantity.
For instance, if we assume $\rho = \epsilon \rho_1$, with $\rho_1 = O(1)$, then Gauss's law becomes: $\nabla \cdot \mathbf{E}_0= \rho_1$. But in this case, we would have an electric field at zeroth order, which is in somewhat at odds with the spirit of a ``magnetic" limit.

More interestingly, we consider the case $\rho = \epsilon^2 \rho_2$, with $\rho_2 = O(1)$. The only difference with respect to Eqs. (\ref{order1_mag1})-(\ref{order1_mag2}) is that Gauss's law now reads as: $\nabla\cdot \mathbf{E}_1 = \rho_2$. In summary, the equations at zeroth and first orders become, expressed in terms of the fields:
\begin{eqnarray}
\nabla \cdot \mathbf{E}_1 &=& \rho_2,  \label{order01_mag_mod1}\\
\nabla \cdot \mathbf{B}_0 &=& 0, \label{order01_mag_mod2} \\
\nabla\times \mathbf{B}_0 &=&  \mathbf{J}, \label{order01_mag_mod3}\\
\nabla\times \mathbf{E}_1 &=& -\frac{\partial \mathbf{B}_0}{\partial t}. \label{order01_mag_mod4}
\end{eqnarray}
The above equations are now identical to those of LBLL for the magnetic limit. It will appear later that the condition $\rho = \epsilon^2 \rho_2$ yields the correct Lorentz transformations of the four-current in the magnetic limit (see Sec. \ref{sec:lorentz}).

We also point out that it was already recognized in some textbooks on classical electromagnetism \cite{Jackson} that the presence of free charges ($\rho \neq 0$) in a model based on the magnetic limit is a second-order effect in $\beta$. This effect is usually neglected in standard MHD, which assumes that $\rho =0$.

Using Eq. (\ref{efield1}) and the Coulomb gauge condition $\nabla \cdot \mathbf{A}_0=0$, we can write Eqs. (\ref{order01_mag_mod1})-(\ref{order01_mag_mod4}) in terms of the potentials:
\begin{eqnarray}
\Delta \mathbf{A}_0 &=& -\mathbf{J} \\
\Delta \phi_1 &=& -\rho_2. \label{first_magnet}
\end{eqnarray}
Notice that the above equations are the exact analog of the equations (\ref{elect_lim_pot_a})-(\ref{elect_lim_pot_b}) that were obtained in the electric limit, except that the orders are inverted (electric effects appear at zeroth order in the electric limit, but at first order in the magnetic limit).

\paragraph{Continuity equation.---}
Pushing the expansion to second order, we find that
\be
\nabla \times \mathbf{B}_2 = \frac{\partial \mathbf E_1}{\partial t},
\ee
so that $\partial_t(\nabla \cdot \mathbf E_1)=\partial_t\rho_2=0$. This is compatible with the condition $\nabla\cdot \mathbf J=0$. In summary, the continuity equation can be written as follows:
\be
\partial_t \rho_2+ \nabla\cdot \mathbf J=0,
\ee
where each term is zero.

\paragraph{Lorentz force.---}
Using Eq. (\ref{lorentz_force}) and remembering that, in the magnetic limit, $\mathbf{E}_0=\mathbf{B}_1=0$ and $\rho=\epsilon^2 \rho_2$, we obtain:
\be
\delta \mathbf F= \epsilon^3 \rho_2 \mathbf{E}_1 +  \epsilon\mathbf{J} \times \mathbf{B}_0.
\ee
Thus, there is no force at zeroth order and only a magnetic force at first order. The electric force appears at third order and is thus uninfluential.

\paragraph{Condition of validity.---}
Using Eq. (\ref{ratio_EB}) and considering again that, in the magnetic limit, $E_0=B_1=0$ while $B_0 \sim E_1=O(1)$, we obtain:
\be
\frac{|\mathbf{E}|}{|\mathbf{B}|} \approx c~\frac{\epsilon E_1}{B_0} \approx c \epsilon
\ee
and since $\epsilon \ll 1$, we have that $|\mathbf{E}| \ll c|\mathbf{B}|$.
This is the condition obtained by LBLL for the validity of the magnetic limit.

\section{Gauge relations}
Let us begin with the general Lorentz gauge condition
\be
\frac{1}{c^2} \frac{\partial\phi}{\partial t} + \nabla \cdot \mathbf A=0.
\ee
Using our dimensionless variables, this becomes
\be
\epsilon \frac{\partial\phi}{\partial t} + \nabla \cdot\mathbf A=0.
\ee

Now, in the electric limit we have $\phi_1=\mathbf{A}_0=0$, so that the gauge relation becomes
\be
\frac{\partial\phi_0}{\partial t} + \nabla \cdot \mathbf A_1=0,
\ee
i.e., {\em in the electric limit the natural gauge is the Lorentz gauge.}

In the magnetic limit we have $\phi_0=\mathbf{A}_1=0$, so that the gauge relation becomes
\be
\epsilon^2 \frac{\partial\phi_1}{\partial t} + \nabla \cdot \mathbf A_0=0.
\ee
At first order in $\epsilon$, the first term in the above equation can be neglected, so that we are left with: $\nabla\cdot \mathbf{A}_0=0$,
i.e., {\em in the magnetic limit the natural gauge is the Coulomb gauge.}

Once again, these ``natural" gauge conditions, well known in the literature, arise automatically when applying our expansion procedure.

\section{Lorentz transformations}\label{sec:lorentz}
In this section, we apply the above expansion technique to the Lorentz transformations of the space-time coordinates, the electric and magnetic fields, and the sources. We will obtain -- in a simple and systematic way -- all the known results in both limits.

\subsection{Space-time}\label{sec:spacetime}
Under a Lorentz transformation, the space-time 4-vector $(ct,\mathbf x)$ transforms as follows:
\begin{eqnarray}
{\mathbf x}' &=& {\mathbf x}-\gamma \mathbf{v}t+(\gamma-1)\frac{\mathbf{v}(\mathbf{v} \cdot \mathbf{x})}{v^2}, \\
t' &=& \gamma\left(t - \frac{\mathbf{v} \cdot \mathbf{x}}{c^2} \right),
\label{lorentz_xt}
\end{eqnarray}
where $\mathbf{v}$ is the relative velocity between the unprimed and primed reference frames.
For low velocities, the Lorentz factor $\gamma$ goes like $\gamma \sim 1-\beta^2/2$ so that at first order in $\epsilon=\beta$ we can take $\gamma=1$. This yields:
\begin{eqnarray}
{\mathbf x}' &=& {\mathbf x}- \mathbf{v}t, \\
t' &=& t - \frac{\mathbf{v} \cdot \mathbf{x}}{c^2}.
\end{eqnarray}

Now we normalize time to $T$, space to $L$, and velocity to $V=L/T$.
We obtain
\begin{eqnarray}
{\mathbf x}' &=& {\mathbf x}- \mathbf{v}t, \\
t' &=& t - \epsilon^2 \mathbf{v} \cdot \mathbf{x}.
\label{galilei_xt}
\end{eqnarray}
To second order in $\epsilon$, the Eq. (\ref{galilei_xt}) becomes simply $t'=t$, thus yielding the standard Galilean transformations.

Note that, since $|\mathbf x| \sim L$, $t\sim T$, we have that: $|\mathbf x|/(ct) \sim L/(cT) \sim \epsilon$. This yields $|\mathbf x| \ll ct$, so that the 4-vector $(ct,\mathbf{x})$ is ultra-timelike.

\subsection{Electric and magnetic fields}\label{sec:fields}
The Lorentz transformations for the fields are as follows:
\begin{eqnarray}
{\mathbf E}' &=& \gamma ({\mathbf E} + \mathbf{v} \times {\mathbf B})+(1-\gamma)\frac{\mathbf{v}(\mathbf{v} \cdot {\mathbf E})}{v^2}, \\
{\mathbf B}' &=& \gamma ({\mathbf B} - {1 \over c^2}~\mathbf{v} \times {\mathbf E})+(1-\gamma)~\frac{\mathbf{v}(\mathbf{v}\cdot {\mathbf B})}{v^2}.
\label{lorentz_eb}
\end{eqnarray}
With the approximation $\gamma \simeq 1$ (valid up to second order) and  using the normalized units defined in Sec. \ref{sec:intro}, one obtains ($\beta=\epsilon$):
\begin{eqnarray}
{\mathbf E}' &=&  {\mathbf E} + \epsilon~ \mathbf{v} \times {\mathbf B}, \label{lorentz_eb1_1}\\
{\mathbf B}' &=&  {\mathbf B} - \epsilon~ \mathbf{v} \times {\mathbf E}.
\label{lorentz_eb1_2}
\end{eqnarray}

Now we expand the electric and magnetic fields in the usual way:
${\mathbf E}={\mathbf E}_0 + \epsilon {\mathbf E}_1 +\dots$, ${\mathbf B}={\mathbf B}_0 + \epsilon {\mathbf B}_1 +\dots$. Substituting into Eqs. (\ref{lorentz_eb1_1})-(\ref{lorentz_eb1_2}) and matching order by order yields:
\begin{eqnarray}
{\mathbf E}'_0 &=&  {\mathbf E}_0 \label{galilei1}\\
{\mathbf B}'_0 &=&  {\mathbf B}_0 \label{galilei2} \\
{\mathbf E}'_1 &=&  {\mathbf E}_1 + \mathbf{v} \times {\mathbf B}_0, \label{galilei3} \\
{\mathbf B}'_1 &=&  {\mathbf B}_1 - \mathbf{v} \times {\mathbf E}_0.  \label{galilei4}
\end{eqnarray}

In the electric limit, $\mathbf{E}_1=\mathbf{B}_0=0$, therefore the Eqs. (\ref{galilei1})-(\ref{galilei4})
reduce to
\begin{eqnarray}
{\mathbf E}'_0 &=&  {\mathbf E}_0 \\
{\mathbf B}'_1 &=&  {\mathbf B}_1 - \mathbf{v} \times {\mathbf E}_0. \label{galilei_e}
\end{eqnarray}
These are the correct Galilean transformations of the fields in the electric limit \cite{LBLL}.

In the magnetic limit, $\mathbf{B}_1=\mathbf{E}_0=0$, therefore the Eqs. (\ref{galilei1})-(\ref{galilei4})
reduce to
\begin{eqnarray}
{\mathbf B}'_0 &=&  {\mathbf B}_0 \\
{\mathbf E}'_1 &=&  {\mathbf E}_1 + \mathbf{v} \times {\mathbf B}_0. \label{galilei_m}
\end{eqnarray}
These are the correct Galilean transformations of the fields in the magnetic limit \cite{LBLL}.

\subsection{Current and density}\label{sec:sources}
The current density 4-vector $(c\rho, \mathbf{J})$  transforms as follows:
\begin{eqnarray}
\mathbf{J}' &=& \mathbf{J}-\gamma  \rho \mathbf{v} +(\gamma-1) \frac{\mathbf{v} (\mathbf{v}  \cdot \mathbf{J})}{v^2}, \\
\rho' &=& \gamma\left(\rho - \frac{\mathbf{v}  \cdot \mathbf{J}}{c^2} \right).
\label{lorentz_jrho}
\end{eqnarray}
Taking $\gamma=1$ and using dimensionless variables, we get
\begin{eqnarray}
\mathbf{J}' &=& \mathbf{J}-\mathbf{v}  \rho, \\
\rho' &=& \rho - \epsilon^2 \mathbf{v}  \cdot \mathbf{J}, \label{lorentz_jrho_nondim}
\end{eqnarray}
which to first order in $\epsilon$ yield finally
\begin{eqnarray}
\mathbf{J}' &=& \mathbf{J}-\mathbf{v} \rho, \\
\rho' &=& \rho.
\end{eqnarray}
The above equations are the standard Galilean transformations in the electric limit. Indeed, since $\rho \sim \overline{\rho}$ and $|\mathbf{J}| \sim \overline{J} = V\overline{\rho}$, one gets that $c\overline{\rho} = \overline{J}/\epsilon \gg |\mathbf{J}|$. The 4-current $(c\rho, \mathbf{J})$ is thus an ultra-timelike vector.

For the magnetic limit, we assume -- as was done in Sec. \ref{sec:magnetic_limit} -- that the density is a second order quantity, i.e., $\rho=\epsilon^2 \rho_2$. Then Eqs. (\ref{lorentz_jrho_nondim}) become:
\begin{eqnarray}
\mathbf{J}' &=& \mathbf{J}-\epsilon^2 \mathbf{v} \rho_2, \label{lorentz_jrho_nondim2}\\
\rho_2' &=& \rho_2 - \mathbf{v} \cdot \mathbf{J}.
\end{eqnarray}
The second order term in Eq. (\ref{lorentz_jrho_nondim2}) can be neglected, so that we finally obtain:
\begin{eqnarray}
\mathbf{J}' &=& \mathbf{J}, \\
\rho_2' &=& \rho_2 - \mathbf{v} \cdot \mathbf{J}.
\end{eqnarray}
This is the standard Galilean transformation of the current and density for the magnetic limit. Indeed, since $|\mathbf{J}| \sim \overline{J} = V\overline{\rho}$ and $\rho \sim \overline{\rho}\epsilon^2$, we have that
\be
\frac{|\mathbf{J}|}{c\rho} \sim \frac{V}{c\epsilon^2}  \sim \frac{1}{\epsilon}
\ee
and thus $|\mathbf{J}| \sim \epsilon^{-1} c\rho \gg c\rho$. In this case, the 4-current is an ultra-spacelike vector.
Note also that in the strictly neutral limit ($\rho=0$), the relation $|\mathbf{J}| \gg c\rho$ is {\it a fortiori} satisfied.

Finally, it is possible to show \cite{Holland-Brown} that the reduced Maxwell equations, both in the electric and in the magnetic limit, are Galilei covariant according to the respective non-relativistic transformations of the space-time coordinates, the electromagnetic fields and the sources, as derived in the preceding paragraphs.

\section{Lagrangian formulation}\label{sec:lagrangian}
In the Lorentz gauge, the Lagrangian density for an electromagnetic field obeying Maxwell's equations can be written in the form \cite{Oosten,Konopinski}:
\be
\mathcal{L}= \frac{\varepsilon_0}{2} (\nabla \phi)^2 - \frac{\varepsilon_0}{2c^2}\left(\frac{\partial \phi}{\partial t}\right)^2
- \frac{(\nabla \mathbf{A})^2}{2\mu_0} + \frac{1}{2\mu_0 c^2}\left (\frac{\partial \mathbf{A}}{\partial t}\right)^2
 -\rho \phi + \mathbf{J} \cdot \mathbf{A}.
\ee
By applying the Euler-Lagrange equations to the above Lagrangian density
\be
\frac{\partial}{\partial t} \frac{\partial \mathcal{L}}{\partial {\dot \psi}}
+ \sum_j \partial_j \frac{\partial \mathcal{L}}{\partial(\partial_j \psi)} -
 \frac{\partial \mathcal{L}}{\partial \psi} =0
\ee
(where $\psi$ is either the scalar potential or one of the components of the vector potential and $\partial_j$ is the $j$'th component of the gradient operator $\nabla$), we obtain the usual wave equations for the scalar and vector potentials in the Lorentz gauge
\begin{eqnarray}
-\Delta \phi + \frac{1}{c^2}\frac{\partial^2 \phi}{\partial t^2} &=& \frac{\rho}{\varepsilon_0},\\
-\Delta \mathbf{A} + \frac{1}{c^2}\frac{\partial^2 \mathbf{A}}{\partial t^2} &=& \mu_0 \mathbf{J}~.
\end{eqnarray}

Let us rewrite the Lagrangian density in the usual dimensionless variables (where $\mathcal{L}$ is normalized to $\overline{\rho}\overline{\phi}$~). We obtain:
\be
\mathcal{L'}= \frac{(\nabla \phi)^2}{2} - \frac{\beta^2}{2} \left(\frac{\partial \phi}{\partial t}\right)^2
- \frac{(\nabla \mathbf{A})^2}{2} + \frac{\beta^2}{2} \left (\frac{\partial \mathbf{A}}{\partial t}\right)^2
 -\frac{\rho \phi}{\alpha} + \frac{\beta}{\alpha}~\mathbf{J} \cdot \mathbf{A},
\ee
where we have defined for convenience $\mathcal{L'}=\mathcal{L}/\alpha$. Since we are interested in developments up to first order, we can disregard the two terms proportional to $\beta^2$. This yields finally:
\be
\mathcal{L'}= \frac{(\nabla \phi)^2 }{2} - \frac{(\nabla \mathbf{A})^2}{2}
 -\frac{\rho \phi}{\alpha} + \frac{\beta}{\alpha}~\mathbf{J} \cdot \mathbf{A},
 \label{lagrangian_firstorder}
\ee

For the electric limit, as usual, $\beta \sim \epsilon \ll 1$ and $\alpha=O(1)$. Expanding the potentials in powers of $\epsilon$, one gets: $\mathcal{L'}=\mathcal{L'}_0+\epsilon\mathcal{L'}_1$, with
\begin{eqnarray}
\mathcal{L'}_0 &=& \frac{(\nabla \phi_0)^2}{2}  - \frac{(\nabla \mathbf{A}_0)^2}{2} -\frac{\rho \phi_0}{\alpha},\\
\mathcal{L'}_1 &=& \nabla \phi_0\cdot \nabla \phi_1 -\nabla \mathbf{A}_0 \cdot \nabla \mathbf{A}_1 -\frac{\rho \phi_1}{\alpha}+ \frac{\mathbf{J} \cdot \mathbf{A_0}}{\alpha}.
\end{eqnarray}
Applying the Euler-Lagrange equations to $\phi_0$, $\mathbf{A}_0$ and their derivatives, we obtain at zeroth order:
\begin{eqnarray}
\Delta \phi_0 &=& -\rho/\alpha, \\
\Delta \mathbf{A_0} &=& 0.
\end{eqnarray}
At first order, the Euler-Lagrange equations with respect to $\phi_1$ and $\mathbf{A}_1$ yield the same equations as above. In contrast, applying the Euler-Lagrange equations to $\phi_0$ and $\mathbf{A}_0$ results in
\begin{eqnarray}
\Delta \phi_1 &=& 0, \\
\Delta \mathbf{A_1} &=& -\mathbf{J}/\alpha.
\end{eqnarray}
If the functions $\phi_1$ and $\mathbf{A_0}$ are zero at infinity, then they must vanish everywhere, so that we recover the usual equations for the electric limit expressed in terms of the potentials.

In the magnetic limit, we have $\beta \sim \alpha \sim \epsilon \ll 1$. If we also take  $\rho=\epsilon^2 \rho_2$, we obtain from Eq. (\ref{lagrangian_firstorder}) at zeroth and first order
\begin{eqnarray}
\mathcal{L'}_0 &=& \frac{(\nabla \phi_0)^2}{2}  - \frac{(\nabla \mathbf{A}_0)^2}{2} + \mathbf{J} \cdot \mathbf{A_0},\\
\mathcal{L'}_1 &=& \nabla \phi_0\cdot \nabla \phi_1 -\nabla \mathbf{A}_0 \cdot \nabla \mathbf{A}_1 -\rho_2 \phi_0+ \mathbf{J} \cdot \mathbf{A_1}.
\end{eqnarray}
Proceeding as was done above for the electric limit, yields the equations:
\begin{eqnarray}
\Delta \mathbf{A_0} &=& -\mathbf{J}, \\
\Delta \phi_1 &=& -\rho_2,
\end{eqnarray}
together with $\Delta \phi_0 = \Delta \mathbf{A_1} =0$, which imply that the latter potentials vanish identically. Once again, we have found the correct equations for the magnetic limit.

\section{Other limits}\label{sec:other}
From the present analysis, it appears that the electric and magnetic limits are the only nontrivial Galilean limits of the Maxwell equations. However, they do not appear to be perfectly symmetric. For the electric limit, no additional assumption had to be made on the sources, whereas for the magnetic limit the charge density must be a second- or higher-order quantity in $\epsilon$. The magnetic limit requires that the system be ``quasi-neutral" (using plasma physics terminology) and therefore both negative and positive charges must be present.

Now, the question is whether it is possible to obtain an electric limit that is also quasi-neutral. In plasma physics, for instance, it is common to employ approximate models that are purely electrostatic (Poisson's equation) and also quasi-neutral. The rationale behind this approximation is that a plasma can be non-neutral only on distances shorter than the Debye length $\lambda_D$. Over longer distances, free charges are screened and the plasma is basically a neutral medium, although electric fields can still be present\footnote{In one of the most popular plasma physics textbooks \cite{Chen} one can read the following statement:  ``In a plasma, it is usually possible to assume $n_e =n_i$ and $\nabla \cdot \mathbf{E} \neq 0$ at the same time. This is
a fundamental trait of plasmas, one which is difficult for the novice to understand". This corresponds to the fact that a very small charge density ($\rho \sim \epsilon$) can still generate a finite electric field.}. Writing Poisson's equation in normalized units: $\Delta \phi = -\rho/\alpha $, it is clear that the charge density must be of order $\alpha=(\lambda_D/L)^2$ in order to generate a finite electric potential.

In the present context, one could obtain such an approximate model by first going to the electric limit ($\beta \to 0$, with $\alpha$ finite) and then taking $\alpha\to 0$. But this procedure is not completely satisfactory, because there is no reason why the relative smallness of the parameters $\alpha$ and $\beta$ should not be specified from the start. In the following paragraphs we suggest a way to derive, in a more rigorous way, an electric limit that is also quasi-neutral.

In deriving the standard electric limit, we had assumed that $\beta \sim \epsilon$ and $\alpha \sim 1$. Lets us now ``raise the order" by one unit, i.e., $\beta \sim \epsilon^2$ and $\alpha \sim \epsilon$, so that we still have $\beta /\alpha \sim \epsilon$. Further, in order to have a zeroth-order electric field, we must require that $\rho = \epsilon \rho_1$, with $\rho_1=O(1)$.
Then, Maxwell's equations for the zeroth and first order fields read as:
\begin{eqnarray}
\nabla\cdot {\mathbf E_0} &=& \rho_1, ~~~~~ \nabla\times {\mathbf B_1} ={\mathbf J},  \label{maxwell_quasi-el1}\\
\nabla\cdot {\mathbf B_1} &=& 0, ~~~~~~
\nabla\times {\mathbf E_0} = 0
\,.
\label{maxwell_quasi-el2}
\end{eqnarray}
together with $\mathbf E_1 = \mathbf B_0 = 0$.
The above equations are sometimes referred to as the ``quasi"static" limit of Maxwell's equations, because they are obtained by neglecting all time derivatives in the original equations.

In terms of the scalar and vector potential and assuming the Coulomb gauge $\nabla \cdot {\mathbf A_1}=0$, these equations can be written as:
\be
\Delta \phi_0 = -\rho_1, ~~~~~~~ \Delta {\mathbf A_1} = -{\mathbf J}.
\ee

From Eq. (\ref{lorentz_force}), the Lorentz force becomes:
$\delta {\mathbf F} = \epsilon \rho_1 {\mathbf E_0} + \epsilon^3 {\mathbf J}\times {\mathbf B_1}$, and the last term can be neglected. Therefore, we end up with a purely electric force and a quasi-neutral system, because $\rho \sim \epsilon \sim \alpha\ll 1$.

One can verify that the Lorentz transformations (up to first order in $\epsilon$) for the space-time variables reduce to the standard Galilei transformations. For the fields, since $\beta=\epsilon^2$, we have: ${\mathbf E}' ={\mathbf E}$ and ${\mathbf B' =\mathbf B}$, and for the sources: $\rho_1'=\rho_1$ and ${\mathbf J}' ={\mathbf J} -\varepsilon {\mathbf v}\rho_1$.

Under the above transformations, Gauss's law in Eqs. (\ref{maxwell_quasi-el1})-(\ref{maxwell_quasi-el2}) is invariant, but Amp\`{e}re's law is not. However, since the resulting force is purely electric, only Gauss's law $\nabla \cdot {\mathbf E_0} = \rho_1$ needs to be taken into account. We further note that the force $\delta {\mathbf F} = \epsilon \rho_1 {\mathbf E_0}$ is also invariant under the same transformations. In summary, the resulting model is indeed purely electrostatic (only ${\mathbf E_0}$ counts) and quasi-neutral (since $\rho \sim \epsilon$).

\section{Conclusion}
In this work, we developed a systematic yet simple approach to the non-relativistic limits of Maxwell's equations. When the latter are rewritten in suitable normalized variables, two dimensionless parameters appear. These parameters represent the typical velocity normalized to the speed of light ($\beta$) and the degree of charge neutrality of the system under consideration ($\alpha$).

The main result of this paper was that the non-relativistic limits of Maxwell's equations can be recovered by letting either or both these parameters go to zero: if $\beta \to 0$ and $\alpha=O(1)$, we recover the {\em electric limit}; in contrast, letting $\beta \to 0$ and $\alpha \to 0$, but keeping $\alpha/\beta=O(1)$, leads to the {\em magnetic limit}.

These results were obtained by expanding the electric and magnetic fields in a power series in the relevant smallness parameter, then matching terms at zeroth and first order. This procedure is both rigorous and systematic, and yields all known results on each of the two limits. Most known properties arise naturally within the present approach: for instance, the Lorentz gauge is shown to be the appropriate choice in the electric limit, whereas the Coulomb gauge is more suited to the magnetic limit.

Our approach revealed some hitherto overlooked subtleties of the theory. For instance, we learnt that: (i) the magnetic and electric fields in the limit equations are actually quantities at different orders in $\epsilon$; (ii) in the electric limit, the Lorentz force has no correction at first order; (iii) in the magnetic limit, the charge density must be at least a second-order quantity (and is usually neglected in MHD applications).
{\correct In addition, we could derive an electric limit that is also quasi-neutral, although this limit is Galilei covariant only at zeroth order, which reduces to Gauss's law.}

Other limits might be found if one also expands the sources in powers of $\beta$. When the Dirac four-current is expanded in powers of $c^{-1}$ (i.e., $\beta$), it displays several correction terms beyond the standard Schr\"{o}dinger expression $\Psi^\dag\Psi$.
For instance, the charge density contains a second-order term, arising from the so-called Darwin correction in the Hamiltonian \cite{Foldy}, which reads as: $\rho_2=\frac{q\hbar^2}{8m^2c^2}\Delta(\Psi^\dag\Psi)$. In general, Galilei covariance will be lost at second order, but the resulting limits are still worth investigating as they are important to understand the coupling of the Maxwell and Dirac equations in semi-relativistic effective field theories \cite{Sulaksono}.

{\correct Finally, as a further development of this work, it will be interesting to perform the same analysis on the generalized Maxwell equations including magnetic charge (magnetic monopoles), which are invariant under duality transformations. The present approach might reveal, in that case, some interesting relationships and symmetries between the two Galilean limits.
}

\end{document}